\documentclass[aps,prb,twocolumn,floatfix]{revtex4}
\usepackage{graphicx}

\begin{document}
\title{Transport properties of nanosystems with conventional and unconventional charge density 
waves}
\author{Marcin Mierzejewski and Maciej M. Ma\'ska} 
\affiliation{
Department of Theoretical Physics, Institute of Physics, 
University of Silesia, 40-007 Katowice,
Poland}

\begin{abstract}

We report a systematic study of transport properties of nanosytems with charge density
waves. 
We demonstrate, how the presence of density waves modifies the current--voltage characteristics.
On the other hand hand, we show that the density waves themselves are strongly affected by the
applied voltage. This self--consistent problem is solved within the formalism of 
the nonequilibrium Green functions.
The conventional charge density waves
occur only for specific, periodically distributed ranges of the voltage. Apart from the
low voltage regime, they are incommensurate and the corresponding wave vectors
decrease discontinuously when the voltage increases.     
\end{abstract}
\pacs \ 73.63.Nm, 73.63.-b, 71.45.Lr
\maketitle
\section{Introduction}

Transport properties of nanosystems are very different from those of macroscopic 
conductors. In particular, nonlinear or even irregular current--voltage characteristics 
seem to be intrinsic properties of these systems. Although, a great number of 
theoretical and experimental results concern the quantum dots, there is an increasing 
interest in the transport properties of nanowires and single 
molecules.\cite{chen,park,donhauser,gittins} The main 
reason for such a tendency is their possible application in the future electronic devices. 
Theoretical analysis of the transport phenomena is difficult due to the coupling between 
a nanosystem and macroscopic leads. As a consequence, Coulomb correlations are usually taken
into account only approximately. 

The spatial confinement, that originates from the geometry of the nanosystems,  may be 
responsible for an inhomogeneity of the charge distribution. Additionally, one may expect 
that phenomena typical for low dimensional correlated systems, e.g., charge density waves 
(CDW), occur as well. The most of the research on the conductance of low dimensional 
systems with CDW correlations focus on the sliding density waves.\cite{markovic,mantel,ringland} 
This transport mechanism sets on for incommensurate 
CDW when the applied voltage exceeds the depining threshold. For a finite 
incommensurate CDW systems it has been shown that the transport properties are controlled 
predominantly by the leads.\cite{oxman}  It has also been shown that the conductance
of a commensurate CDW system  is very different  from that of an incommensurate one.
The theoretical description of incommensurate CDW is similar to a theory of the Luttinger liquid.
The resulting temperature dependence of the conductance is much simpler than that of 
commensurate CDW.\cite{krive} 
These results  lead straightforwardly to a question
abouth the mechanisms that determine the commensurability  of CDW in meso- and nanoscales,
e.g, whether commensurability of the charge 
distribution depends on the bias voltage and the geometry of the nanosystem. Recent 
self--consistent investigations of the molecular chain show that charge distribution 
strongly depends on the applied voltage.\cite{bulka,emberly} This effect has already 
been observed in molecular devices.\cite{chen,collier}
In analogy to this result, one may 
expect that the applied voltage changes also the charge distribution in the CDW system, 
what affects the current--voltage characteristic. 
The current itself can modify the charge distribution as well. 
As a result, we may obtain a system, where small changes of the
applied voltage can, through the modification of the charge distribution, drive the system 
between insulating and metallic states. It is possible, that this tempting feature of 
the CDW  nanosystems could be applied in switching devices.

In the present paper we use
the formalism of nonequilibrium Green functions to
analyze the charge distribution in a nanosystem coupled to leads. We demonstrate that
the applied voltage can induce a transition between commensurate and incommensurate CDW.
We analyze how this transition depends on the geometry of the  nanosystem.
The transport properties determined for conventional CDW are compared with the results obtained
for systems with unconventional density waves.\cite{ucdw} In particular, we analyze density 
waves state with $d$-wave symmetry (DDW) that has intensively been investigated as a possible scenario of 
the pseudogap phase in high--temperature superconductors.\cite{ddw} Such an order has also been 
proposed as a low temperature phase of some quasi--two--dimensional organic 
conductors.\cite{maki}

\section{Model and method}

The Hamiltonian of the system under consideration consists of three parts which describe
the nanosystem itself, macroscopic electrodes and the coupling between the electrodes and
the nanosystem: $H=H_{\rm nano} + H_{\rm el} + H_{\rm nano-el}$.  
The electrodes are modeled by a lattice gas of noninteracting electrons with a wide energy band:
%\begin{equation}
$
H_{\rm el}=\sum_{{\bf k},\sigma,\alpha} \left(\varepsilon_{{\bf k},\alpha} -\mu_{\alpha}\right)
c^\dagger_{{\bf k},\sigma,\alpha} c_{{\bf k},\sigma,\alpha},  
$
%\end{equation}
where $\mu$ denotes the chemical potential and $\alpha \in$ \{L,R\} indicates the left or 
right electrode.  $c^\dagger_{{\bf k},\sigma,\alpha} $ creates an electron with momentum 
${\bf k}$ and spin $\sigma$ in the electrode $\alpha$. At the mean field level the 
Hamiltonian of the nanosystem is given by
\begin{eqnarray}
H_{\rm nano}&=&\sum_{\langle ij\rangle\sigma} \left[ -t_{ij}+
(-1)^i U_{\rm DDW} W_{ij}\right] c^\dagger_{i\sigma}c_{j\sigma} \nonumber \\
&-&U_{\rm CDW}\sum_{i\sigma}
\left(\langle n_{i-\sigma}\rangle -\frac{1}{2}\right)c^\dagger_{i\sigma}c_{i\sigma},
\label{HDDW}
\end{eqnarray}
where 
$c^\dagger_{i\sigma}$ creates an electron with spin $\sigma$  
at site $i$ of the  nanosystem, $n_{i\sigma}=c^\dagger_{i\sigma}c_{i\sigma}$ 
and
$
W_{ij}=(-1)^i 
\langle c^{\dagger}_{i\sigma}c_{j\sigma}
-c^{\dagger}_{j\sigma}c_{i\sigma}\rangle/2
$. The potentials $U_{\rm CDW}$ and $U_{\rm DDW}$ describe the strength of interactions that are
responsible for the formation of conventional and unconventional charge density waves, respectively.\cite{imp1,my1} 
The coupling between the nanosystem and the electrodes is described by:
\begin{equation}
H_{\rm nano-el}=\sum_{{\bf k},i,\alpha,\sigma} \left(
g_{{\bf k},i,\alpha} c^\dagger_{{\bf k},\sigma,\alpha }c_{i\sigma}+{\rm H.c.}\right).
\end{equation} 
A few remarks concerning the validity of the mean field approach are needed at this stage.
It is generally believed that the mean field approximation is inappropriate for
low dimensional systems. However, it has been argued that this approximation may be
applicable at low temperatures for weak interaction and strong coupling between 
nanosystem and electrodes (see Ref. \onlinecite{bulka} 
and the discussion therein). Therefore, we restrict our considerations to the case 
$U_{\rm CDW (DDW)} <4t $ and assume a relatively strong coupling between the electrodes
and the nanosystem.
Additionally, the recent investigations of nanorings
with CDW correlations have shown that the mean field results qualitatively agree
with the exact ones\cite{my2}    
(although the quantitative differences remain significant).
These results concern the properties of persistent currents in a system, where CDW is
pinned by impurities. In the present case, the coupling to electrodes should play
a similar role stabilizing CDW. Consequently, we expect that the mean--field analysis
provides correct qualitative results for the transport currents in a CDW system.

The thermal averages, which occur in the Hamiltonian (\ref{HDDW}) are  calculated self--consistently
using the lesser Keldysh Green functions:
\begin{equation}
\langle c^\dagger_{i\sigma} c_{j\sigma} \rangle
=\frac{1}{2 \pi i} \int {\rm d} \omega \: G^{<}_{j\sigma,i\sigma} (\omega).
\end{equation}
After obtaining convergency, one can calculate the current
flowing through the nanosystem:
\begin{eqnarray}
J&=&\frac{2e}{h} \int {\rm d} \omega \left[f_{\rm L}(\omega)-f_{\rm R}(\omega)\right] \nonumber \\
&& \times
{\rm Tr}\left[ \hat{\Gamma}_{\rm L}(\omega)\hat{G}^r(\omega)
\hat{\Gamma}_{\rm R}(\omega)\hat{G}^a(\omega) \right],
\end{eqnarray}
where $f_{\alpha}(\omega)$ is the Fermi function of electrons in the
electrode $\alpha$.
The elements of the matrix $\hat{\Gamma}_{\rm L(R)}$ are given by
\begin{equation}
\left[\hat{\Gamma}_{\alpha}(\omega)\right]_{ij}= 2 \pi \sum_{{\bf k}}
g^{*}_{{\bf k},i,\alpha} g_{{\bf k},j,\alpha} \delta(\omega-\varepsilon_{{\bf k},\alpha}),
\label{Gamy}
\end{equation}
whereas the matrices $\hat{G}^r(\omega)$ and $\hat{G}^a(\omega)$ consist of
retarded and advanced Keldysh Green functions, respectively. 
For the sake of brevity we do not present the complete set of equations that determines the
Green functions for the mean field Hamiltonian. Instead we refer to Ref. \onlinecite{bulka}
for the details.   

\section{Conventional charge density waves}

We start our investigations with a one-dimensional (1D) nanowire taking into account only 
the nearest--neighbor hopping. The ends of the nanowire are connected to the macroscopic leads. 
The only non--vanishing elements of $\hat{\Gamma}$'s are assumed to be frequency
independent
$\left[\hat{\Gamma}_{\rm L}(\omega)\right]_{11}=\left[\hat{\Gamma}_{\rm R}(\omega)\right]_{NN}=\Gamma$,
where the sites in the chain are enumerated from 1 to $N$.
The difference between the lead's potentials gives the voltage 
applied to the nanosystem $eV=\mu_{\rm L}-\mu_{\rm R}$. 

\begin{figure}
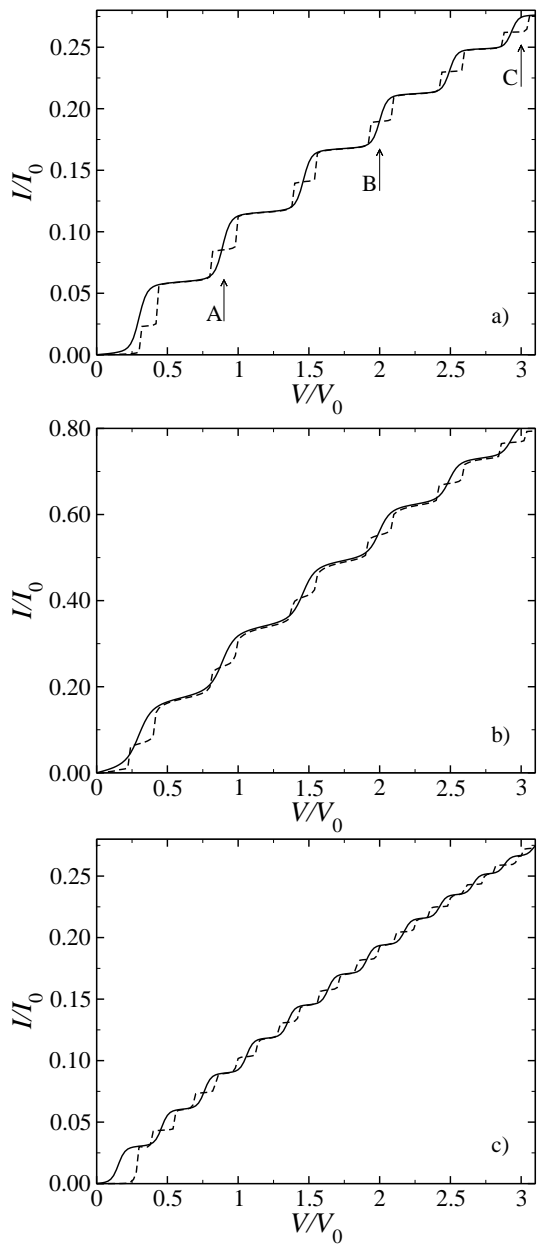

\centerline{\includegraphics[width=7cm]{fig1a.eps}}
\centerline{\includegraphics[width=7cm]{fig1b.eps}}
\centerline{\includegraphics[width=7cm]{fig1c.eps}}
\caption{Current--voltage characteristics of one dimensional chains containing 20 (Figs. 1a and 1b) 
and 40 (Fig. 1c) atoms for the temperature of the leads $k_{B}T=0.01t$. Figs. 1a and 1c
correspond to $\Gamma=0.1t$ and Fig. 1b to $\Gamma=0.3t$. Continuous and dashed lines show 
results obtained for $U_{\rm CDW}=0$ and $U_{\rm CDW}=1.5t$, respectively. We have denoted 
$I_0=2 e t/h$ and $V_0=t/e$, where $t$ in the nearest--neighbor hopping integral.\cite{norm}
Arrows labeled as A, B and C indicate voltages for which the charge distributions
are shown in  Fig. \ref{fig2}.
} 
\label{fig1}
\end{figure}
In a 1D system 
$d$--density wave cannot occur and, therefore, in this case 
we restrict ourselves only to the conventional charge density wave ($U_{\rm DDW}=0$).
Fig. \ref{fig1} shows current--voltage ($I-V$) characteristics of 20-- and 40--atom chains coupled 
to the macroscopic leads. Here, we compare the characteristics obtained for $U_{\rm CDW}=1.5t$ with 
the results for a non interacting system ($U_{\rm CDW}=0$)

\begin{figure}
\centerline{\includegraphics[width=7cm]{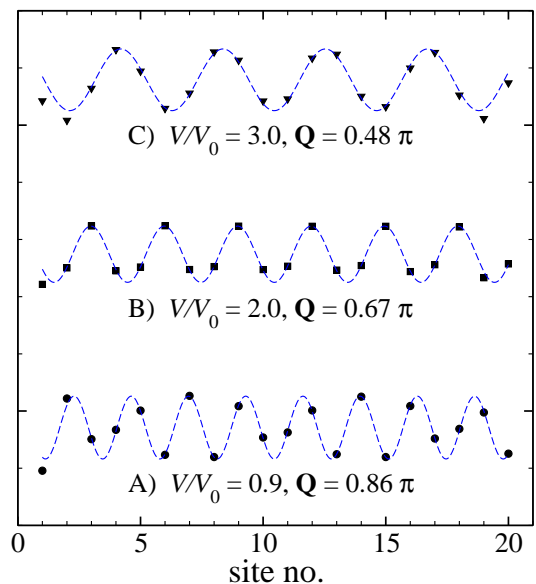}}
\caption {(Color online) Average occupation 
 $\langle n_{i \sigma} \rangle $ obtained for a 20--atom chain with $U_{\rm CDW}=1.5t$
The voltage is indicated explicitly in the figure, whereas the remaining model parameters are the same as in
Fig. \ref{fig1}. Results have been fitted by the function 
$\langle n_{i \sigma} \rangle = n^0+A \cos\left({\bf Q} \cdot {\bf R}_i+ \phi\right) $. 
For clarity of the figure the curves are offset.} 
\label{fig2}
\end{figure}
\begin{figure}
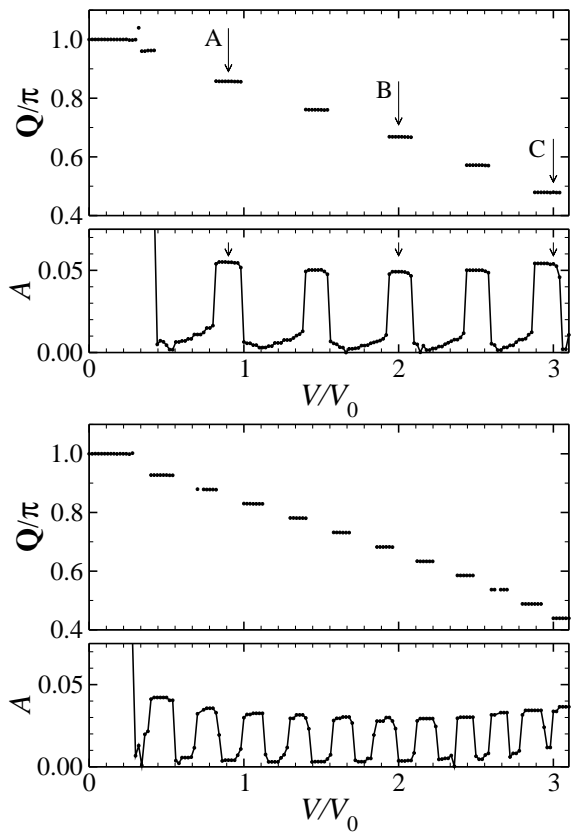

\centerline{\includegraphics[width=7.5cm]{fig3a.eps}}
\centerline{\includegraphics[width=7.5cm]{fig3b.eps}}
\caption{Voltage dependence of the density wave amplitude $A$ and the wave vector $\bf Q $  
for chains containing 20 (upper panel) and 40 
(lower panel) atoms. Model parameters are the same as in Fig. \ref{fig1}. Arrows labeled as
A, B and C denote cases presented in Fig. \ref{fig2}. }\label{fig3}
 \end{figure}

In the uncorrelated case the $I-V$ characteristic consists of a series of plateaus 
smoothly 
connected by steep sections. In the following, we refer to these plateaus as {\em original plateaus}.
In the presence of the CDW interaction additional smaller 
plateaus occur instead of these steep sections and the resulting $I-V$ characteristic
changes from a relatively smooth one to a step function.
Within the original plateaus the characteristic remains almost unchanged.
These results clearly show that the CDW correlations modify the $I-V$
characteristic only for some particular values of $V$. 
It occurs as a result of a strong suppression of the charge density waves
by the applied voltage in the regions of the 
original plateaus. In order to confirm this statement we have analyzed the
spatial distribution of electrons over the nanowire. Fig. \ref{fig2} 
shows the occupation number as a function of position for a 20--atom chain at various voltages. 
We have found that within the original plateaus electrons are distributed almost uniformly over
the system. Contrary to this, in the regions where the additional plateaus occur the 
electron density is strongly inhomogeneous and can be described as (in most cases
incommensurate) charge density waves, i.e,
$\langle n_{i \sigma} \rangle$ can be fitted very accurately by 
$ n^0+A \cos\left({\bf Q} \cdot {\bf R}_i+ \phi\right) $.
 It turns out that the CDW wave vector ${\bf Q}$ strongly 
depends on $V$. For $V=0$ the CDW is commensurate with the lattice ($|{\bf Q}|=\pi$). The same holds 
true for a small voltage. However, when the voltage increases, the CDW wavelength increases as 
well. Fig. \ref{fig3} shows how the density wave vector ${\bf Q}$ and 
and the density wave amplitude $A$ change with the applied voltage. Similarly to the $I-V$ 
characteristic ${\bf Q}(V)$ is a step function, whereas $A(V)$ is approximately
a two value periodic function.
Comparing Figs. \ref{fig1} and \ref{fig3} one can
see that each of the additional plateaus in the $I-V$ characteristic corresponds 
to a different value of  the CDW wave vector. 
The overall behavior of the $I-V$ and ${\bf Q}(V)$ characteristics
obtained for the 20--atom chain is similar to those obtained for the 40--atom wire.
The main difference is related to the number and length of the plateaus. Namely, the number of 
the allowed values of ${\bf Q}$ increases with the increasing length of the nanowire. Moreover, 
the ratio of the lengths of the original and CDW--induces plateaus decreases when the length
of the nanowire increases. It suggests that for a sufficiently long nanowire ${\bf Q}$ should become a
continuous decreasing function of $V$. Comparison of Figs. \ref{fig3}a and \ref{fig3}b illustrates
this tendency.
In an isolated 1D system the CDW wave vector can be changed by the modification of the
Fermi wave vector ($|{\bf Q}|=2k_{\rm F}$), that in turn is a sigle--valued function of
the occupation number.
In the present case  ${\bf Q}$ can be changed independently of the concentration of electrons 
by means of the applied voltage. 

\begin{figure}
\centerline{\includegraphics[width=7.5cm]{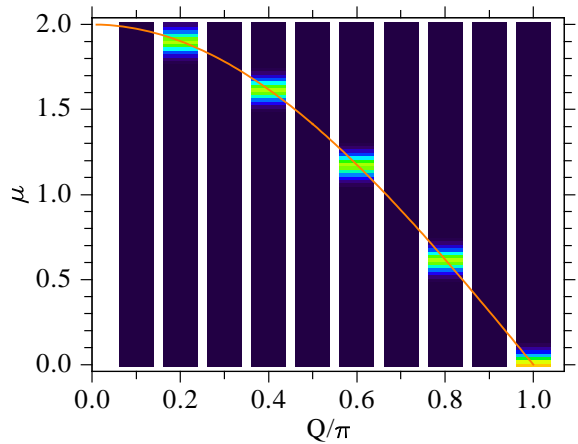}}
%\centerline{\includegraphics[width=8.5cm]{FIG_X/plot.eps}}
\caption{(Color online) CDW susceptibility as a function of the
chemical potential $\mu$ and the wave vector ${\bf Q}$. Lighter regions
correspond to larger values of the susceptibility. These results 
have been obtained for an isolated 20--atom chain with the nearest
neighbor hopping and periodic boundary conditions. The solid line
shows $\mu=2t \cos(|{\bf Q}|/2)$ (see text for explanation).
}\label{fig_new}
\end{figure}

In order to get insight into the physical origin of the above results, one can consider
an isolated chain. In this case, the onset of the
charge density waves is determined by the CDW susceptibility, defined as 
a retarder equilibrium Green function:
\begin{equation}
\chi({\bf Q},\omega)=-\langle\langle \hat{\Delta}({\bf Q}) \mid  \hat{\Delta}^\dagger({\bf Q}) 
\rangle\rangle, 
\end{equation}
where
\begin{equation}
\hat{\Delta}({\bf Q})=\frac{1}{N} \sum_{{\bf k},\sigma} c^{\dagger}_{{\bf k}+{\bf Q},\sigma}
c_{{\bf k},\sigma},
\end{equation}
and $ c^{\dagger}_{{\bf k},\sigma}$ creates an electron with momentum ${\bf k}$ and spin
$\sigma$. In the static case, the CDW susceptibility is given by the Lindhard function:
\begin{equation}
\chi({\bf Q},\omega=0)=\frac{2}{N}\sum_{\bf k} \frac{f\left(\epsilon_{{\bf k}+{\bf Q}}\right)
-f\left(\epsilon_{\bf k}\right)}{ \epsilon_{{\bf k}+{\bf Q}}-\epsilon_{\bf k}},
\label{podat}
\end{equation}
where $\epsilon_{\bf k}=2t \cos(|{\bf k}|)-\mu$ 
%denotes the kinetic energy of an electron with momentum ${\bf k}$,
and $f(\epsilon)$ is the Fermi distribution function. 
Fig. \ref{fig_new} shows $\chi({\bf Q},\omega=0) $ as a function of the chemical
potential $\mu$ for a finite system.   
The maxima of the
CDW susceptibility occur for such ${\bf Q}$'s, that both the energies in the
denominator in Eq.(\ref{podat}) vanish, i.e., for  $\epsilon_{\bf k}=0$  and
${\bf Q}=2 {\bf k}$. In the case of an infinite 
system ${\bf k}$ changes continuously and the first equation has a solution for
an arbitrary value of the chemical potential. 
Then, the maximum of the CDW susceptibility occurs
for ${\bf Q}$ that fulfills the condition $\mu=2t \cos(|{\bf Q}|/2)$.   
On the other hand, for a finite 
system, ${\bf k}$ takes on discrete values and the maxima of the CDW susceptibility
occur only for specific values of $\mu$, i.e., when  $\mu$ is equal to one of the
energy levels. This feature is responsible for the CDW--induced plateaus
in the $I-V$ characteristics. In a case of noninteracting electrons the steps arise
due to resonant tunneling through a multilevel quantum system. The steep sections occur
when the successive energy levels are taking part in the charge transport. 
However, simultaneously the discussed above criterion for the onset of CDW is fulfilled. As a result, the CDW gap opens and new plateaus occur in the middle of these steep sections.  

But still there is a question concerning the degree of steepness of the sections that connect the plateaus.
It determines the height of peaks in the differential conductance. The width
of the one particle energy--levels is related to $\Gamma$. When the system is weakly coupled
to the electrodes the energy levels are very narrow and the $I-V$ characteristic consists
of sharp steps, provided the temperature is low enough. Increasing of $\Gamma$ smooths out these
steps. It holds true both in the presence and in the absence of CDW,  
what can be inferred from Figs. \ref{fig1}a and  \ref{fig1}b.

\begin{figure}
\centerline{\includegraphics[width=7cm]{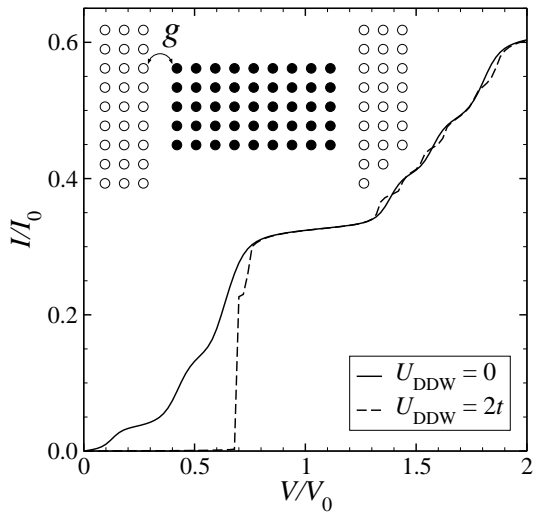}}
\caption{ Current--voltage characteristics of a 10 $\times$ 4 nanosystems coupled to
the leads, as shown in the inset. 
Continuous and dashed lines show results obtained for $U_{\rm CDW}=0$
and $U_{\rm CDW}=2t$, respectively. We have neglected the energy dependence of
$\hat{\Gamma}_{\alpha}(\omega)$ and adjusted 
the value of the  hopping energy $g$ 
in such a way that  
$\left[\hat{\Gamma}_{\alpha}(\omega)\right]_{ii}=0.1 t \delta_{i \alpha}$ 
(see Eq. \ref{Gamy} for the details).}\label{fig4}
\end{figure}

In the following we investigate the influence of the transverse dimension of
the nanosystem. For that purpose we consider a nanowire of a finite width. 
In such a case the description of the 
coupling between the nanosystem and the leads becomes nontrivial. We assume 
a simple model in which leads are described by a two--dimensional (2D) lattice gas and the 
hopping between the leads and nanosystem is possible only perpendicularly to the edge
of the nanosystem (see the inset in Fig. \ref{fig4}). Then, the nonvanishing elements 
of the matrices $\hat{\Gamma}_{\rm L}(\omega)$ and $\hat{\Gamma}_{\rm R}(\omega)$ can
be calculated directly from the Eq. (\ref{Gamy}):
\begin{equation}
\left[\hat{\Gamma}_{\alpha}(\omega)\right]_{ij}= 2 \pi \sum_{{\bf k}}
|g|^2 \delta_{i \alpha} \delta_{j \alpha} \cos\left({\bf k}\cdot {\bf R}_{ij}\right)  
\delta(\omega-\varepsilon_{{\bf k},\alpha}),
\label{Gamy1}
\end{equation}
where $g$ denotes the hoping amplitude and ${\bf R}_{ij}={\bf R}_i-{\bf R}_j$. 
$\delta_{i {\rm R(L)}}$ is equal to 1 if the site $i$ is
located at the right (left) edge of the nanosystem and vanishes otherwise.

Fig. \ref{fig4} shows the $I-V$ characteristics for a 4-site wide and 10-site long
nanosystem calculated for $U_{\rm CDW} =2t$ and $U_{\rm CDW} = 0$. Contrary to the 1D
case an important difference between the correlated and uncorrelated case is visible only
for low voltage. We have found that in this case there exist a commensurate CDW with  
the wave vector ${\bf Q}=(\pi,\pi)$. Increasing of $V$ leads to a disappearance 
of the CDW ordering.  There exist minor differences between both the characteristics 
for larger values of $V$.
However, they appear irregularly and are much smaller than in the 1D case. 
Therefore, the voltage--induced incommensurate CDW seems to be an intrinsic feature 
only of the 1D systems. One may attribute this behavior to general properties of the
density waves systems. Namely, in the 2D case the nesting of the Fermi surface plays
a crucial role for stability of the CDW phase. The CDW wave vector connects the
nested parts of the Fermi surface. 
For a square lattice with the nearest neighbor hopping
the Fermi surface is perfectly nested only in the half--filled case, what corresponds
to the commensurate ${\bf Q}=(\pi,\pi)$. Other values of ${\bf Q}$ do not correspond 
to perfectly nested Fermi surface. 
Therefore, incommensurate CDW in 2D systems is usually less stable than in 1D cases.     
We believe that this property is responsible for the visibly different $I-V$ characteristics
of 1D and 2D systems.
        
\section{Unconventional charge density waves}

\begin{figure}
\centerline{\includegraphics[width=7cm]{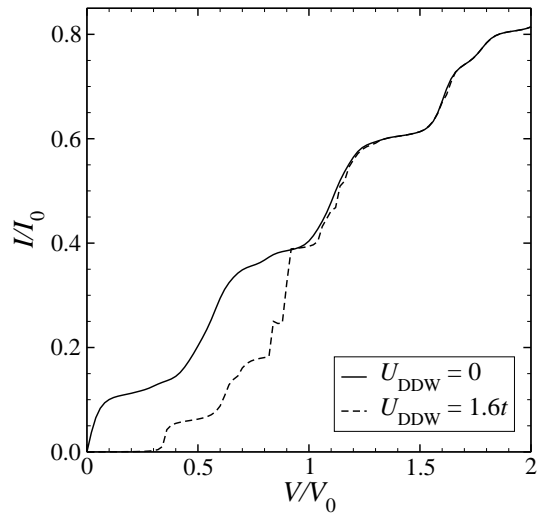}}
\caption{Current--voltage characteristics of a 20 $\times$ 6 nanosystems coupled to
the leads, as shown in the inset of Fig. \ref{fig4}. 
Continuous and dashed lines show results obtained for $U_{\rm DDW}=0$
and $U_{\rm DDW}=1.6t$, respectively. The remaining model parameters 
are the same described in the caption of  Fig. \ref{fig4}.}\label{fig5}
\end{figure}
\begin{figure}
\centerline{\includegraphics[width=6.4cm]{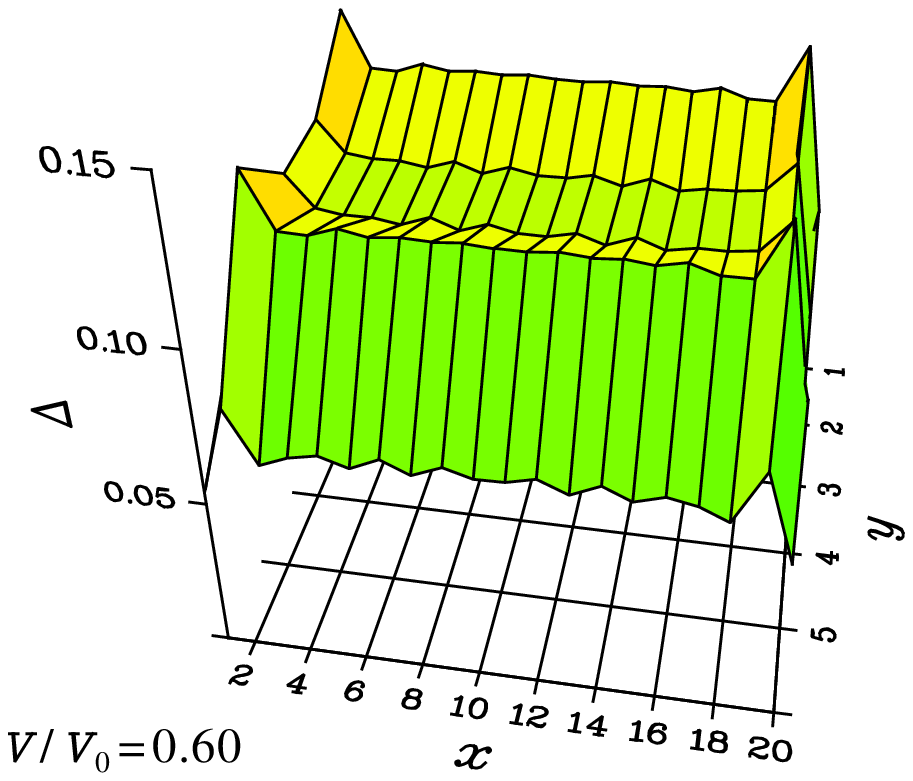}}
\centerline{\includegraphics[width=6.4cm]{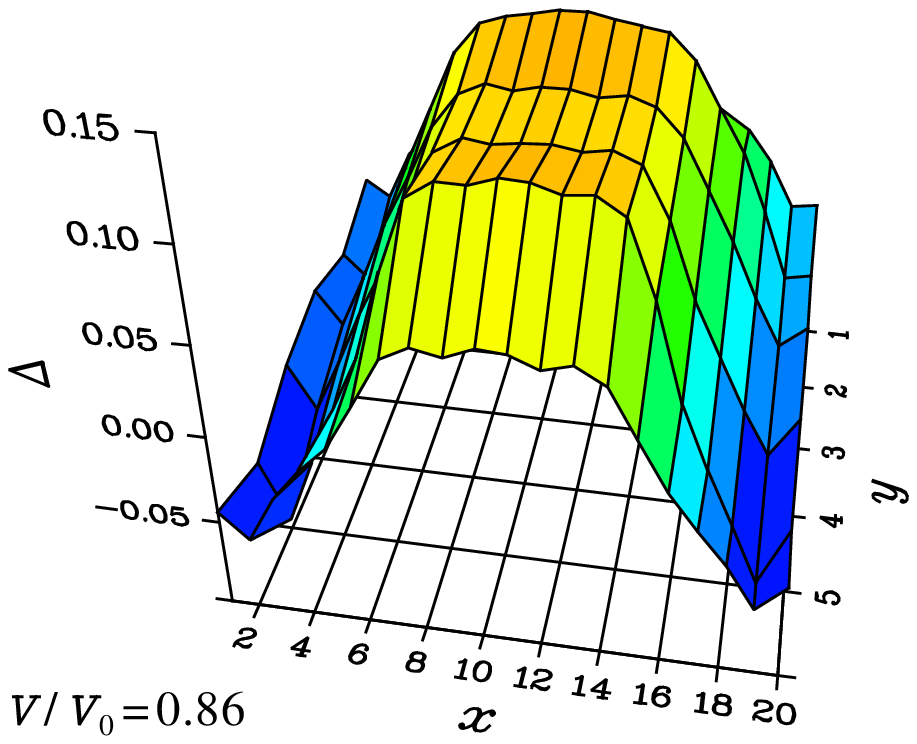}}
\centerline{\includegraphics[width=6.4cm]{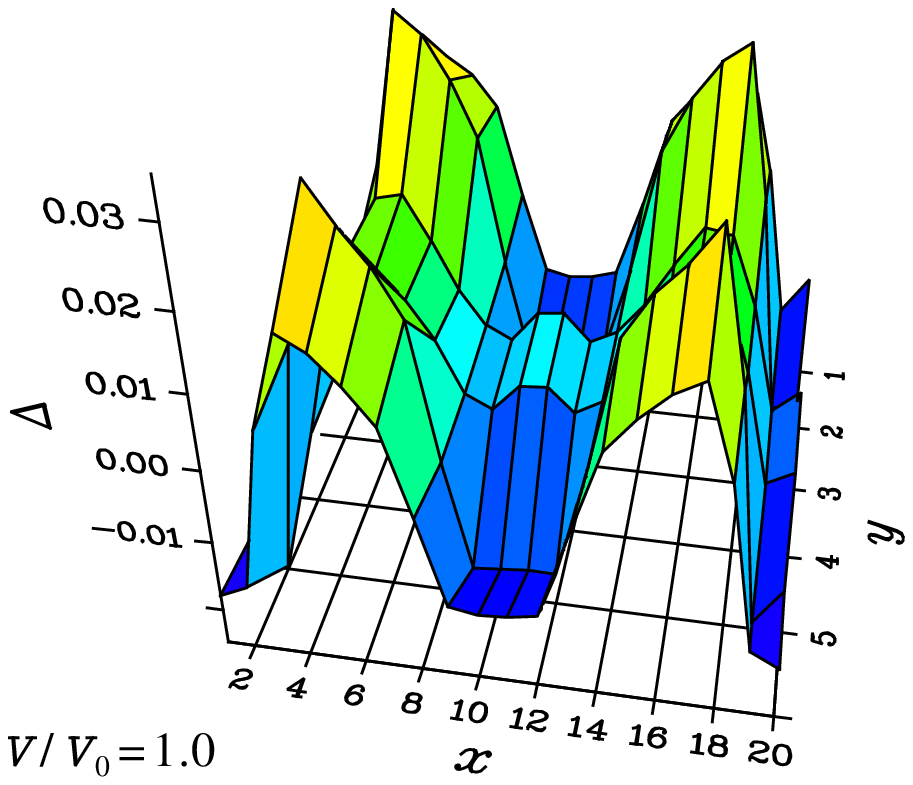}}
\centerline{\includegraphics[width=6.4cm]{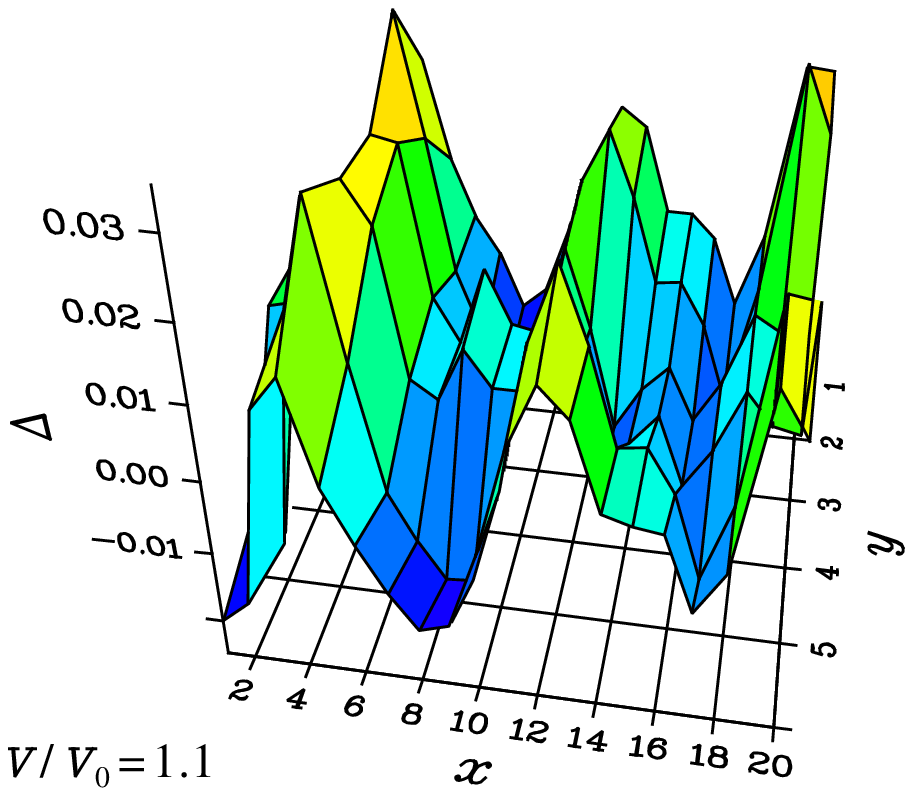}}
\caption{(Color online) Spatial distribution of the orbital currents $\Delta_i$.
The voltage is indicated explicitly in the figure, 
whereas the remaining model parameters are the same as in
Fig. \ref{fig5}. }\label{fig6}
\end{figure}

In the following we extend our analysis by taking into account unconventional density 
waves, where a condensation of electron--hole pairs with non--zero angular momentum 
occurs.\cite{ucdw}
%In the following we extend our analysis taking into account unconventional density 
%waves, origintaing from the formation of electron--hole pairs with non--zero angular 
%momentum.\cite{ucdw}
Such a state, with the angular momentum $l=2$, has recently been proposed 
as an explanation for the pseudogap phenomena in high--temperature superconductors.\cite{ddw} 
Contrary to conventional density waves the charge is distributed uniformly over the whole system,
but there occur orbital currents, i.e., the state breaks the time-reversal symmetry. 
This difference may be visible in the transport properties because of 
the interference between the transport and orbital currents. 
Recent developments in the fabrication techniques allow one to produce nanowires out of
high--temperature superconductors.\cite{bonetti} The pseudogap is visible in these systems. 
Moreover, a discrete switching noise in the resistance of the nanowires has been observed 
in the pseudogap regime and explained in terms of the formation of the stripe phase.  
It suggests that interesting phenomena emerge in 
high--temperature superconductors, when one enters the meso-- and nanoscales.
One may also expect that these new results may contribute to understanding of the pseudogap 
phenomenon. In particular, a question arises, whether 
the results presented in Ref. \onlinecite{bonetti} can be explained within the
unconventional density wave scenario of the pseudogap. 

In order to investigate this problem within the formalism introduced above, we have
considered a system described by the Hamiltonian (\ref{HDDW}) with $U_{\rm CDW}=0$.
Figure \ref{fig5} shows the resulting current--voltage characteristics
obtained for $U_{\rm DDW}=1.6t$ and $U_{\rm DDW}=0$. They are very similar
to those of conventional 2D CDW for low and sufficiently high voltages.
In the first case, i.e, for low voltage the energy gap does not allow for the current flow and the
system is insulating. In the opposite case, the applied voltage destroys both the 
conventional and unconventional density waves. However, for the unconventional density waves there exists also
an intermediate regime, where the transport current is finite but the $I-V$ characteristic
significantly differers from the results obtained for the uncorrelated system ($U_{\rm DDW}=0$).
In this regime sharp steps in the $I-V$ characteristic occur,
indicating on the rapid changes of the orbital currents distribution.
Fig. \ref{fig6} shows the spatial distributions of the orbital currents $\Delta_i$ at various 
voltages, where $\Delta_i$ is given by:
\begin{equation}
\Delta_i=\frac{1}{4}\left(
W_{i,i+{\bf\hat x}}+W_{i,i-{\bf\hat x}}
-W_{i,i+{\bf\hat y}}-W_{i,i-{\bf\hat y}}
\right).
\label{deldef}
\end{equation}

For a low voltage $\Delta_i$ is almost independent of the
lattice site $i$, what  indicates on a uniform magnitude of the orbital currents. 
The sudden drop of $\Delta_i$ at the system edges originates from the reduced number 
of the neighboring sites [see Eq. (\ref{deldef})]. Increase of $V$ reduces
the magnitude of the orbital currents. Additionally, this quantity becomes spatially modulated, as
can be inferred from Fig. \ref{fig6}.
The modulation is mostly visible in the longitudal direction. It is an almost periodic modulation, with the 
period decreasing with increasing $V$. This behavior is opposite to the previously discussed 1D conventional CDW, 
where the period of the charge modulation increases with $V$.  
There exist lines where  $\Delta_i$  changes sign, what corresponds 
to reverted circulation of the orbital currents. 
The voltage induced transitions 
between various distributions of $\Delta_i$ are accompanied by sharp steps in the $I-V$
characteristic. Therefore, one could speculate that in larger systems such transitions may be responsible for
the switching noise in the resistance of the nanowires in the pseudogap phase,\cite{bonetti}  i.e., 
in the phase, that could be described as unconventional density waves state.\cite{ddw}

\section{Discussion and concluding remarks}

In order to investigate the transport properties of nanosystems with charge density waves
we have applied the formalism of nonequilibrium Keldysh Green functions.
Both the conventional and unconventional states have been considered.
The most of the already published results concern the sliding CDW, 
where it is a priori assumed whether the density waves
are commensurate or not. It has previously been shown that the transport
properties of commensurate and incommensurate CDW systems are different.
On the other hand, it is known that commensurability of an isolated 1D CDW system 
depends on the position of the Fermi level, or equivalently on the occupation number. 
In the case of transport through the nanosystem, its properties
are determined by the chemical potentials of the left and right electrodes, which are shifted 
by the applied voltage. We have shown that in the case of 1D CDW system  
the commensurability changes with  the voltage, whereas the average concentration 
of electrons remains unchanged. These CDW states
occur only for specific, periodically distributed ranges of the voltage. The
number of the allowed values of the ${\bf Q}$ vector increases with the increase of the length of 
the system. Therefore, we expect that for a sufficiently long nanowire ${\bf Q}$ should 
linearly decrease with the applied voltage.  
We have shown that the applied voltage affects also the unconventional density waves. 
In particular, it leads to spatial modulation of the orbital currents, especially in the
longitudal direction. It is a remnant of the voltage--dependent charge modulation, 
that occurs in a 1D system with conventional CDW. However, contrary to the CDW case the period of this
modulation decreases with the increase of $V$. The difference between the $I-V$ characteristics
obtained for 2D nanosystem with conventional and unconventional CDW
is attributed to the interference between the transport and orbital currents that occurs in the
latter case.

To summarize, we have shown that the properties of nanosystems with conventional
and unconventional density waves strongly depend on the applied voltage. The discussed mechanism
should be taken into account also in the analysis based on the sliding CDW mechanism, 
since the transport properties depend on the commensurability.

\acknowledgments
This work has been supported by 
the Polish Ministry of Education and Science under Grant No. 1~P03B~071~30.


\begin{thebibliography}{99}
\bibitem{norm}  The transport current is often expressed in units of 
$2 e \Gamma /h $. In the following, we investigate also a 2D case, when
the coupling between the electrodes and nanosystem is described by the matrix 
$\hat{\Gamma}_{ij}$. In order to use the same units in 1D and 2D cases, 
instead of $\Gamma$ we make use of the hopping integral $t$.
\end{thebibliography}
\end{document}